\documentclass[a4paper,11pt]{article}
\usepackage{jinstpub} % for details on the use of the package, please see the JINST-author-manual
\usepackage{lineno}
\usepackage{multirow}
\usepackage{booktabs}

\title{The CRILIN calorimeter: gamma radiation resistance of crystals and SiPMs}

\author[a]{A.~Cemmi,}
\author[a]{B.~D'Orsi,}
\author[b]{E.~Di~Meco,}
\author[a]{I.~Di~Sarcina,}
\author[b,1]{E.~Diociaiuti,\note{Corresponding authors.}}
\author[b]{M.~Moulson,}
\author[b]{D.~Paesani,}
\author[b]{I.~Sarra,}
\author[a]{J.~Scifo}
\author[a,1]{and A.~Verna}

\affiliation[a]{ENEA Nuclear Department, NUC-IRAD-GAM Laboratory\\Via Anguillarese, 301, I-00123, Roma, Italy}
\affiliation[b]{INFN, Laboratorio Nazionale di Frascati\\Via E. Fermi, 54, I-00044, Frascati (RM), Italy}

\emailAdd{eleonora.diociaiuti@lnf.infn.it, adriano.verna@enea.it}
\abstract{
The Crilin calorimeter is a semi-homogeneous calorimetric system based on Lead Fluoride (PbF$_2$) crystals with UV-extended Silicon Photomultipliers (SiPMs)  proposed for the Muon Collider. This study investigates the radiation resistance of crystals and SiPMs, subjected to 10 kGy gamma irradiation, equivalent to a 10-year service life in the Muon Collider.

Our findings indicate that while PbF\(_2\) crystals exhibit a decrease in transmittance post-irradiation with partial recovery over time, the alternative PbWO\(_4\)-Ultra Fast (PWO-UF) demonstrates exceptional radiation hardness, maintaining stable transmittance. SiPMs showed an increase in dark current and breakdown voltage post-irradiation, with less degradation observed in the SiPM biased during the exposure to radiation compared to the unbiased component.

These results underscore the viability of PbF\(_2\) for radiation-tolerant calorimeters, though improvements in production homogeneity are needed. 
The superior performance of PWO-UF crystals suggests they are a promising alternative for high-radiation applications, but their higher cost must be carefully considered.}


\begin{document}
\maketitle
\flushbottom

\section{Introduction}
Innovative accelerator technologies must be developed to explore the energy frontier and advance future high-energy physics research. In this context, the Muon Collider proposal is particularly promising. As thoroughly explained in ref.~\cite{accettura2023towards}, a Muon Collider offers several advantages over hadron-based accelerators: it can achieve higher energies than electron-based accelerators due to lower Bremsstrahlung losses, and it has a limited hadronic background with a precisely known initial state free from parton density function uncertainties. However, implementing this plan presents several challenges.

One of the main challenges from a detector perspective is Beam Induced Background (BIB), characterized by an intense radiation environment caused by muon decay products interacting with accelerator components, generating secondary and tertiary particles. Montecarlo simulations predict for the electromagnetic calorimeters a 1~MeV neutron equivalent (1-MeV-neq) fluence of about 10$^{14}$~cm$^{-2}\cdot$y$^{-1}$ and a total ionizing dose (TID) of $\sim$1~kGy$\cdot$y$^{-1}$, as shown in figures 20 and 21 of ref.~\cite{accettura2023towards}, respectively. To mitigate this, optimizing the machine-detector interface is essential. 

The current solution for the Muon Collider involves 64 million readout channels for the electromagnetic calorimeter barrel, achieved by placing $5 \times 5$ mm$^2$ silicon sensors in 40 layers of tungsten. This configuration is expensive and technologically challenging, prompting the investigation of alternative strategies. A notable option is CRILIN (CRystal calorImeter with Longitudinal INformation), a semi-homogeneous calorimeter based on lead fluoride (PbF$_2$). Each crystal is read out by two series of UV-extended surface-mounted Silicon Photomultipliers (SiPMs). More information about the CRILIN calorimeter status and features can be found in refs.~\cite{Cemmi_2022, Ceravolo_2022, NIMA, Frontiers, BTF}. In addition to the baseline solution of the PbF$_2$, another very promising crystal, the PWO$_4$-Ultra Fast (PWO-UF), was studied. It is obtained by doping the PbWO$_4$ matrix with lanthanum and yttrium atoms that act as recombination centers, resulting in a very fast luminescence response. Nevertheless, the radiation hardness of this new material has been tested only up to a dose of 150 Gy \cite{KORZHIK_22}. Testing the radiation hardness of the calorimeter’s main components, including crystals and SiPMs, is crucial for verifying its effectiveness and reliability in high-radiation environments. Although the radiation hardness of PbF$_2$ and SiPMs was verified at high gamma absorbed doses in a few previous papers \cite{Achenbach_98,Ren_GH_14,Cemmi_2022,Cantone_23a,BTF, GARUTTI201969}, the present work focuses on the behavior of CRILIN's most important single components (crystals and SiPMs)  after gamma irradiation at roughly 10 kGy absorbed dose that corresponds to a 10-year service life of the Muon Collider.  The fast recovery of the possible radiation-induced damage is also studied. The effect of neutrons and the investigation of the radiation damage on the entire calorimeter prototype will be the subject of forthcoming publications.

\section{The Cherenkov and Emission Detection Weights: crystals and SiPMs coupling}

A good crystal candidate for CRILIN must preserve its optical properties under irradiation but also fulfill many other requirements: high density (i.e., high stopping power), fast emission response, reproducibility of growth conditions, absence of neutron activation, and moderate cost. On the contrary, a high light yield is not as necessary due to the very high energy (hundreds of GeV) of the particles to be detected.

To assess radiation effects on a detection system using PbF$_2$ and PWO-UF crystals, we must consider both the Photon Detection Efficiency (PDE) function of the SiPMs \cite{HamamatsuData} and the spectral distribution of radiation emitted by the crystals. For PbF$_2$, a pure Cherenkov radiator, the spectral distribution of Cherenkov radiation was used \cite{Fienberg_15,Frankenthal_19}. For PWO-UF, both Cherenkov radiation and photoluminescence (PL) were considered. Given that the measured total light yield of pristine PWO-UF crystals is approximately double that of PbF$_2$ \cite{Frontiers}, and assuming an equal Cherenkov yield for both materials, we can infer that Cherenkov and PL contribute roughly equally to the emission from PWO-UF crystals.

\begin{figure}[ht]
\centering
\includegraphics[width=\textwidth]{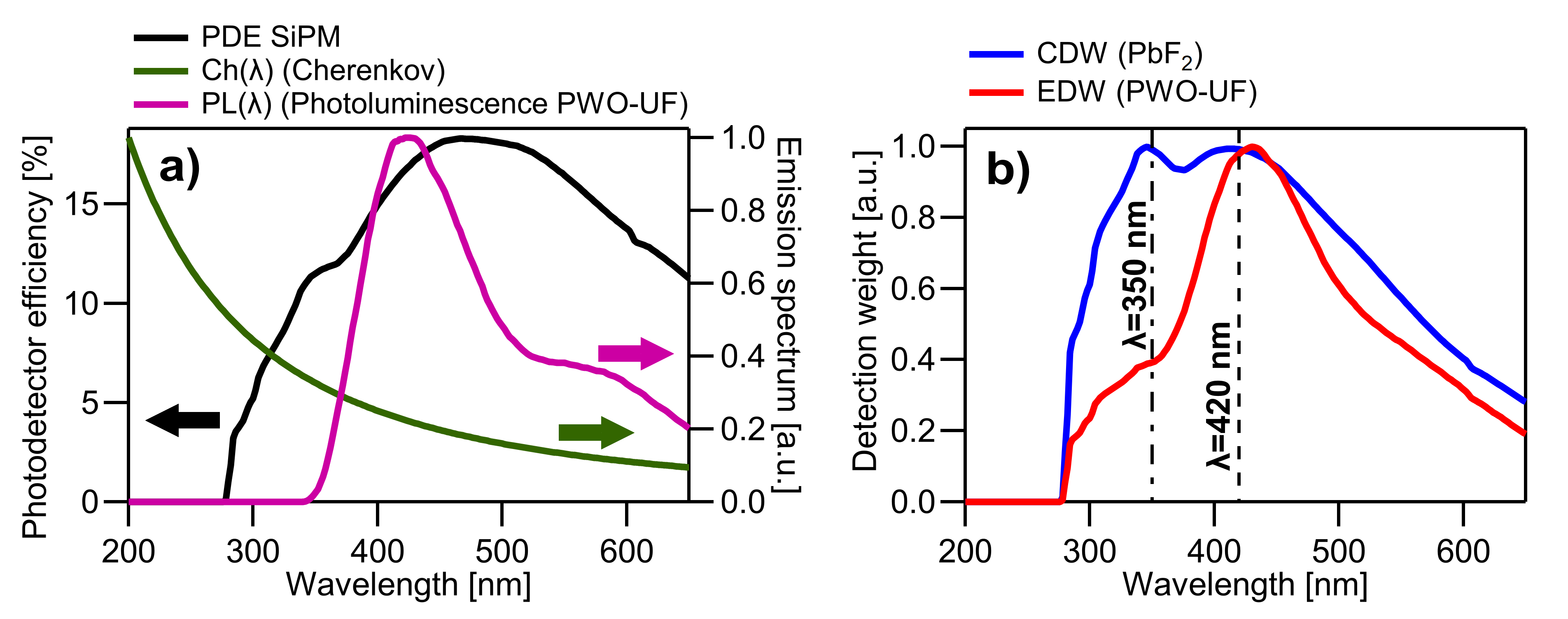}
\caption{(a) Photodetection efficiency (PDE) of the SiPMs (black curve), the approximate spectral distribution of the Cherenkov photons $\mbox{Ch}(\lambda)\propto 1/\lambda^2$ (blue curve), and the photoluminescence spectrum $\mbox{PL}(\lambda)$ of PWO-UF as obtained from ref.~\cite{KORZHIK_22} (magenta curve). (b) CDW (blue curve) and EDW (red curve) functions. The vertical lines indicate the $\lambda$=350~nm (dash-dotted) and $\lambda$=420~nm (dashed) wavelength values.}
\label{fig:CDW}
\end{figure}

The spectrum of Cherenkov photons $\mbox{Ch}(\lambda)$, which roughly decreases as 1/$\lambda^2$, is shown in figure~\ref{fig:CDW}a along with the photoluminescence spectrum $\mbox{PL}(\lambda)$ of PWO-UF, obtained via optical excitation \cite{KORZHIK_22}, and the PDE of the SiPMs. The Cherenkov Detection Weight (CDW) function, which is the product of PDE and $\mbox{Ch}(\lambda)$, is presented in figure~\ref{fig:CDW}b. This function provides insight into the spectral regions contributing to Cherenkov radiation detection by the SiPMs, and thus identifies the wavelength range for which high transmittance is critical in the case of PbF$_2$ crystals. Similarly, for PWO-UF crystals, we define an Emission Detection Weight (EDW) as the product of PDE  by the sum of the $\mbox{Ch}(\lambda)$ and $\mbox{PL}(\lambda)$ functions, weighted so that they provide the same contribution for $\lambda$>$\lambda_\textrm{cutoff}\approx$350~nm (figure~\ref{fig:CDW}b). CDW and EDW curves in figure~\ref{fig:CDW}b are normalized to their maximum to qualitatively highlight the spectral regions of interest.

As shown in figure~\ref{fig:CDW}b, the CDW function presents two maxima around $\lambda$=350~nm and $\lambda$=420~nm, respectively. These two maximum positions are used as reference wavelength values to monitor transmittance variations induced by gamma irradiation for PbF$_2$. The wavelength $\lambda$=420~nm is also very close to the peak of EDW fuction (430~nm) and can be conveniently used as the reference to evaluate transmission loss in PWO-UF, which is completely opaque at 350~nm even in the pristine form.

\section{The Calliope irradiation facility}
\begin{figure}[ht]
\centering
\includegraphics[width=0.8\textwidth]{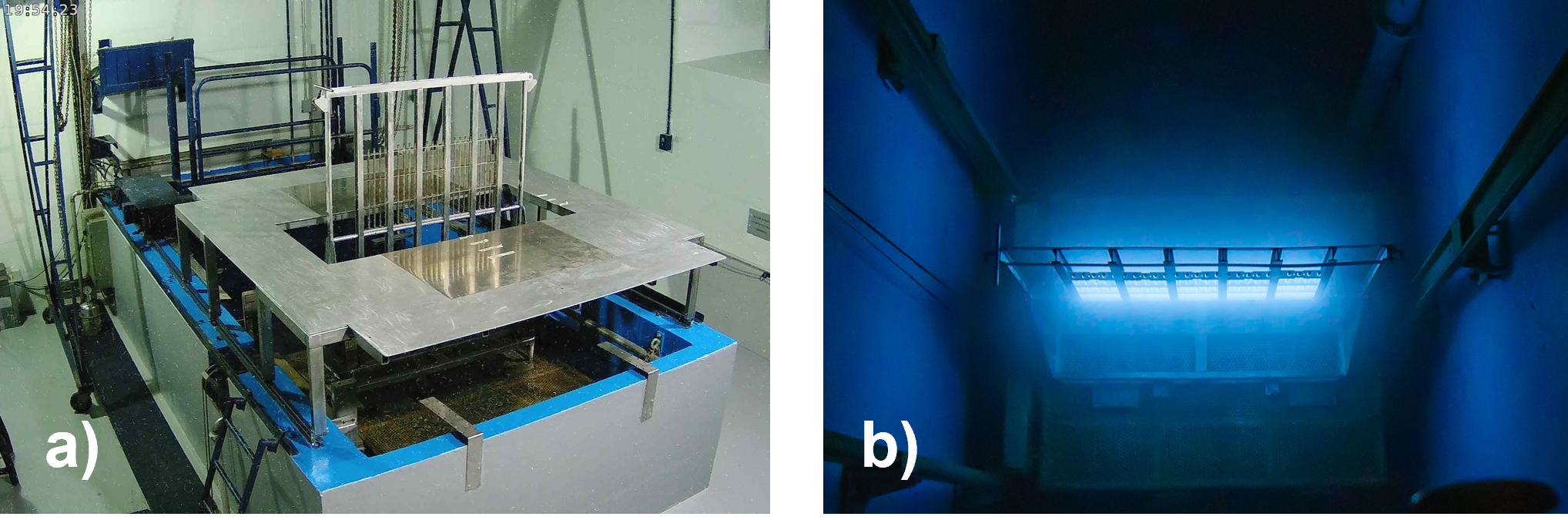}
\caption{Image of the irradiation cell with the source rack in the irradiation position (a) and  on the bottom of the pool in the recovery position (b).}
\label{fig:Calliope}
\end{figure}

Gamma irradiation of crystals and SiPMs was carried out at the Calliope irradiation facility of the ENEA Casaccia Research Center (Rome, Italy) \cite{Calliope_report}. Calliope is a pool-type irradiation plant equipped with a $^{60}$Co $\gamma$ source (mean energy 1.25~MeV) in a high volume (7.0~m $\times$ 6.0~m $\times$ 3.9~m) shielded cell. The source consists of 25 cylindrical rods of $^{60}$Co mounted on a rack with plane geometry (active area of 41~cm $\times$ 75~cm). Photos of the irradiation facility are shown in figure~\ref{fig:Calliope}.

The dose rate used in this study was measured through the alanine-EPR dosimetry system \cite{ISO_51607_12} with a dose rate and absorbed dose uncertainty of 5~\%.

During the irradiation, the crystals were placed 20 cm from the $^{60}$Co source with the longer axes of the crystals perpendicular to the plane of the source rack, to reproduce the working conditions of the crystals in the calorimeter, with particles impinging from only one side. The crystals were irradiated at 0.7 kGy and 7.2 kGy absorbed doses (referred to air) with a mean dose rate of 2.9~kGy/h (estimated in the center of the crystals).

Regarding SiPMs irradiation, the two samples of the first set were placed in front of the source parallel to the surface of the rack during the exposure and were irradiated at dose rate of 3.3~kGy/h and at absorbed dose of 10~kGy (referred to silicon). 

\section{Optical properties of crystals before and after gamma irradiation}
Inorganic scintillators are typically made of a crystal or glass matrix where luminescence originates either from intrinsic emission centers already present in the structure or from doping agents/impurities intentionally introduced into the crystal lattice \cite{Rodnyi1997, Weber2004}. Unlike scintillators, which rely on luminescence, Cherenkov crystals emit radiation when a charged particle travels through them at a speed greater than the phase velocity of light in that medium \cite{Potylitsyn2021}.
When exposed to ionizing radiation, such as gamma rays or neutrons, both crystal lattice and impurities can be altered, affecting the optical properties of the system. For instance, ionizing radiation can displace electrons from their lattice positions creating electron-hole pairs that can become trapped by charged defects, forming color centers \cite{Lin2001, Deng1999, Wensheng2000}. These color centers absorb light in the ultraviolet–visible (UV–VIS) range \cite{Baccaro2015, Baccaro2001} reducing the crystals' optical transmission \cite{Addesa2020}, and resulting in a lower quantity of Cherenkov and/or scintillation light reaching the photodetector, with negative impact on the detection efficiency.

In this work, two PbF$_2$ monocrystals and one PWO-UF monocrystal, doped with La and Y at a total level of 1500 parts per billion, were investigated. The PbF$_2$ crystals, labeled C1 and C2, were produced by SICCAS\texttrademark{} in the same batch and grown in the cubic $\beta$ phase with no intentional doping with the Bridgman method, while the PWO-UF crystal was grown by Crytur\texttrademark{} with the Czochralski method. The three crystals are parallelepipeds with dimensions $10 \times 10 \times 40 \, \text{mm}^3$ and optical lapping on all surfaces. To prevent photobleaching effects, the crystals were kept in the dark during the entire experiment (irradiation, optical measurements and recovery study) \cite{Achenbach_98}.

The optical properties of the investigated samples were studied by comparing their transmittance (\%T) curves recorded with a PerkinElmer\texttrademark{} Lambda 950 UV-VIS-IR double-beam spectrophotometer, equipped with an integrating sphere, in the range 200-650 nm with a 2 nm step. Longitudinal transmittance measurements were performed immediately before and after the irradiation steps. To evaluate the possible transmittance recovery, additional measurements were conducted 1 hour after the end of irradiation at the 7.2~kGy absorbed dose. The error on transmittance measurements was $\pm$2\%.

\begin{figure}[ht]
\centering
\includegraphics[width=0.5\textwidth]{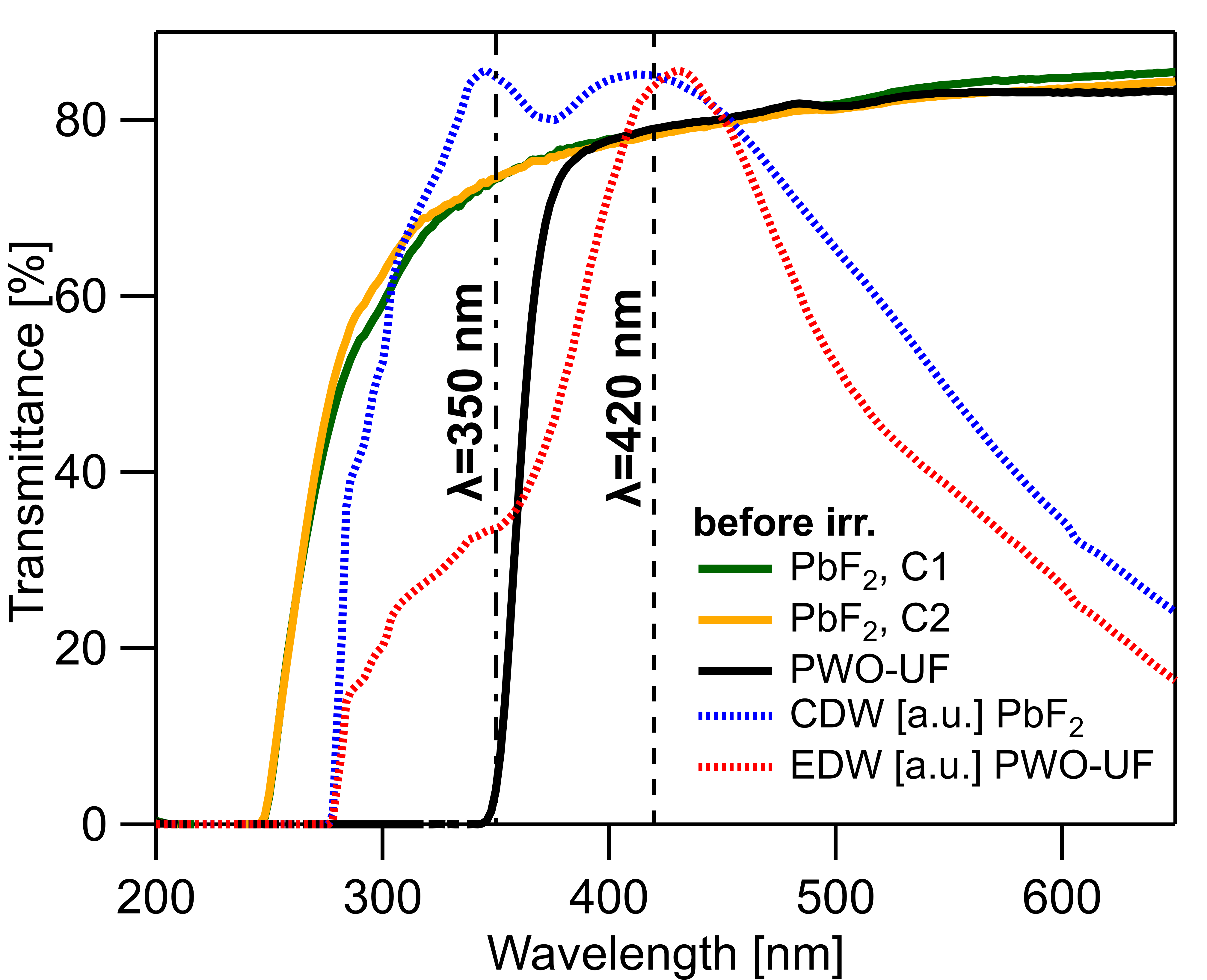}
\caption{Transmittance spectra acquired before the irradiation of C1 (green solid curve), C2 (orange solid curve), and PWO-UF (black solid curve) crystals superimposed with the CDW function (blue dotted curve) and the EDW function (red dotted curve). The vertical lines indicate the $\lambda=350$~nm (dash-dotted) and $\lambda=420$~nm (dashed) wavelength values.}
\label{fig:transmittanceBef}
\end{figure}

The transmittance spectra of the crystals before irradiation, together with the CDW and EDW functions, are reported in figure~\ref{fig:transmittanceBef}. 

As expected, PbF$_2$ crystals show a transmittance cutoff around 248 nm, whereas PWO-UF has an absorption edge around 351 nm. As previously discussed, since SiPMs can detect radiation with wavelengths greater than 275~nm, the PWO-UF cuts a significant portion of the detectable radiation, whereas the PbF$_2$ transmission region completely contains the range of SiPMs operation.

\begin{figure}[ht]
\centering
\includegraphics[width=12 cm]{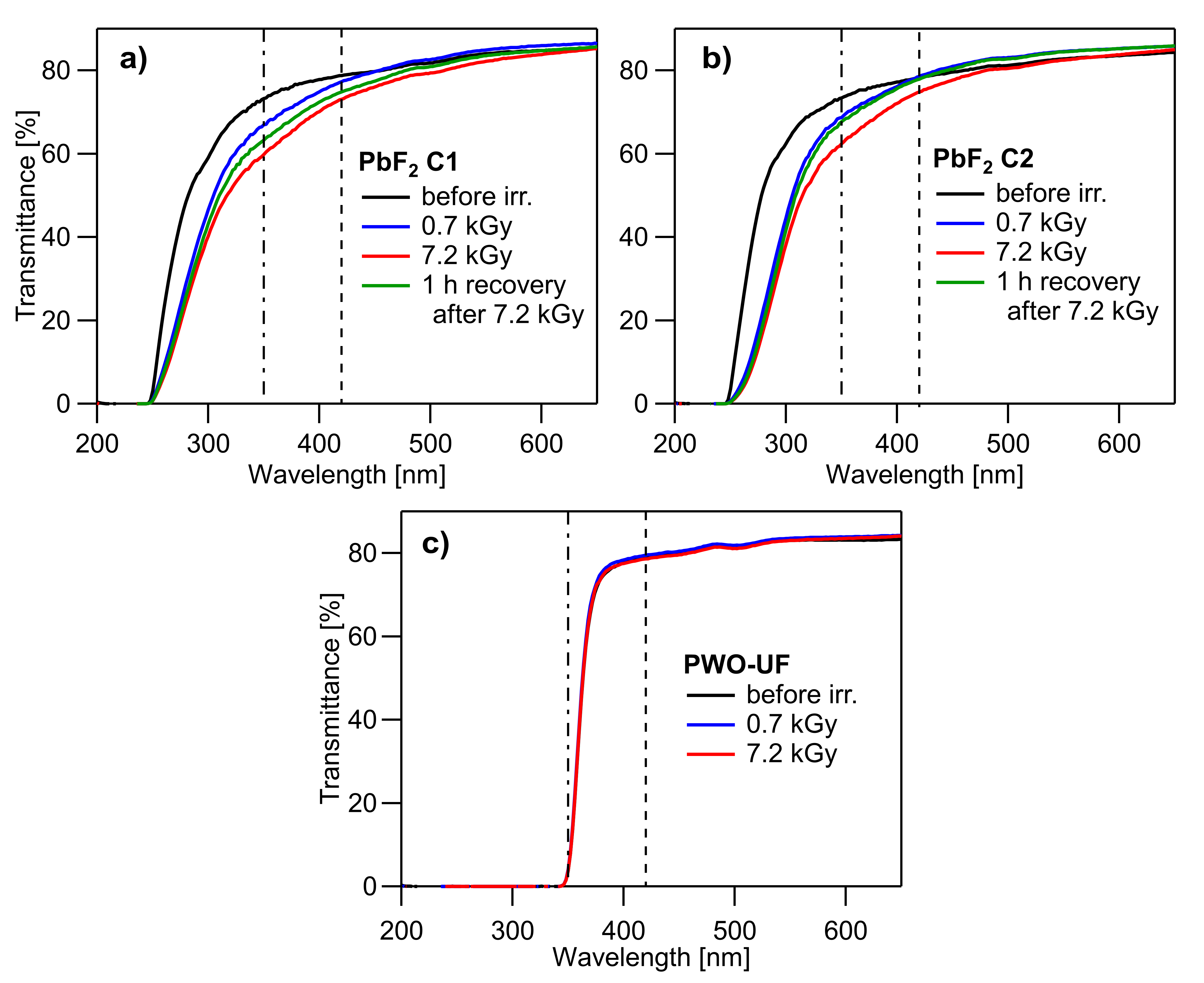}
\caption{Transmittance spectra measured before irradiation (black curve), after 0.7 kGy (blue curves), after 7.2 kGy (red curve), after 7.2 kGy with $\sim$1 hour recovery (green curve) for the C1 (a), C2 (b), and PWO-UF (c) crystals. For the PWO-UF crystal there is only one acquisition after the second irradiation step. In the legend the cumulative absorbed doses are indicated. The vertical lines indicate the $\lambda=350$~nm (dash-dotted) and $\lambda=420$~nm (dashed) wavelength values.}
\label{fig:dopoIrr}
\end{figure}

The transmittance spectra after gamma irradiation are reported in figure~\ref{fig:dopoIrr} for PbF$_2$ C1 (a), PbF$_2$ C2 (b), and PWO-UF (c). The spectra acquired before irradiation are also shown for comparison, as well as those acquired for PbF$_2$ crystals 1 hour after the end of irradiation at 7.2 kGy to study their fast recovery. 

PbF$_2$ crystals present a decrease in transmittance at $\lambda$=350~nm even with 0.7~kGy absorbed dose. Increasing the absorbed dose at 7.2~kGy determines a further loss of transparency that is in this case significant also at $\lambda$=420~nm. One hour after the end of irradiation to 7.2 kGy, a slight recovery was observed, although differently from each PbF$_2$ crystal. In particular, the recovery is more pronounced in C2 sample, for which it is evident at both the reference wavelengths, while in C1 crystal an appreciable recovery is found only at $\lambda$=350~nm. On the contrary, the PWO-UF crystal shows strong radiation hardness with no measurable loss of transmittance at all the investigated absorbed doses. 
The \%T values at $\lambda$=350~nm and 420~nm for all the samples at the different stages are summarized in Table~\ref{tab:transmittance}.

\begin{table}[ht]
\centering
\begin{tabular}{|l|l l|l l|l|}
\hline
&\multicolumn{5}{c|}{\textbf{Transmittance [\%T]}}\\
\hline
&\multicolumn{2}{c|}{\textbf{PbF$_2$ C1}} &\multicolumn{2}{c|}{\textbf{PbF$_2$ C2}} & \textbf{PWO-UF} \\
\hline
 & 350 ~nm & 420~nm & 350 ~nm & 420~nm & 420 nm \\ 
\hline
 Before irradiation & 73.3 & 78.7 & 73.4 & 78.3  & 79.0 \\ \hline
 0.7 kGy & 67.0 & 77.3 & 68.7 & 78.6 & 79.5 \\ \hline
 7.2 kGy & 59.9 & 73.1 & 62.5 & 74.9 & 78.6 \\ \hline
 1 hour recovery after 7.2 kGy & 63.3 & 74.8 & 67.7 & 77.9 & —\\
\hline 
\end{tabular}
\caption{Measured transmittance at $\lambda$=350~nm and 420~nm for the PbF$_2$ crystals, and at $\lambda$=420~nm for the PWO-UF crystal. The error associated with the transmittance measurement is $\pm 2 \%$.}
\label{tab:transmittance}
\end{table}

The cutoff wavelength $\lambda_{cutoff}$, calculated as the intercept of the linear fit of the optical absorption edge with the $\%T=0$ axis, results 248$\pm$1~nm for both the pristine PbF$_2$ crystals. Irradiation induces a red shift of $\lambda_{cutoff}$ to 250$\pm$1~nm and to 256$\pm$1~nm for the C1 and C2 samples, respectively, after an absorbed dose of 0.7~kGy. After irradiation at 7.2~kGy absorbed dose, a further shift to longer wavelength (252$\pm$1~nm and 260$\pm$1~nm for C1 and C2, respectively) was also observed.  This red shift with increasing absorbed doses indicates the formation of negatively charged defects (electron centers, ECs), absorbing in the UV range. No changes in term of cutoff are observed after recovery. For the PWO-UF crystal the cutoff wavelength remains stable at 351$\pm$1~nm also after irradiation, supporting the evidence of a greater radiation resistance for this material.

In conclusion, despite the transmittance curves of pristine PbF$_2$ crystals are almost identical, after gamma irradiation and recovery different results are shown for C1 and C2 samples in term of transmittance and cutoff. 
This different behavior for two crystals of the same batch indicates that the quality production in terms of homogeneity is not sufficiently good and hence should be improved.

\section{Irradiation of SiPMs}

Discussing the effect of radiation on silicon detectors, it is important to differentiate between two types of damage: (i) bulk damage, which results from Non-Ionizing Energy Loss (NIEL) and is measured by Displacement Damage (DD), and (ii) surface damage, which results from Ionizing Energy Loss (IEL) and is quantified by the Total Ionizing Dose (TID) \cite{GARUTTI201969,LINDSTROM}.
 In this article, the effects of gamma radiation on pivotal characteristics such as the SiPM dark current ($I_{\text{dark}}$) and operational voltage are reported. Specifically, two 10~$\mu$m pixel Hamamatsu\texttrademark{} 3 $\times$ 3 mm$^2$ MPPC\textsuperscript{\textregistered} (multi-pixel photon counter) samples of the type no.~S14160-3010PS \cite{HamamatsuData} were investigated.

Both SiPMs were read and biased using a Keithley 6487 Picoammeter \cite{PicoAmmeter} controlled remotely through a Python graphic interface. The temperature was set at 25$^{\circ}$C, and the scans were performed in the range 32V-46V with a step of 300 mV.
To determine the breakdown voltage of the SiPMs, a fit to the current-voltage (I-V) curves has been performed. The function used \cite{para} is reported in the following:
\begin{equation}
    I(V_{bias})=\begin{cases}
    (I_0 \times V_{bias})+C\times (1-e^{-p\cdot(V_{bias}-V_{br})})\times(V_{bias}-V_{br}) , & V>V_{br}\\
    I_0\times V_{bias} , & \text{otherwise}.
  \end{cases}
\end{equation}
where $V_{bias}$ is the bias voltage, $V_{br}$ the breakdown voltage, $I_0$ the current before the breakdown, $P$ the Geiger probability, and $C$ is proportional to the number of free carriers (both thermal and optical).

\begin{figure}[ht]
    \centering
     \includegraphics[width= \textwidth]{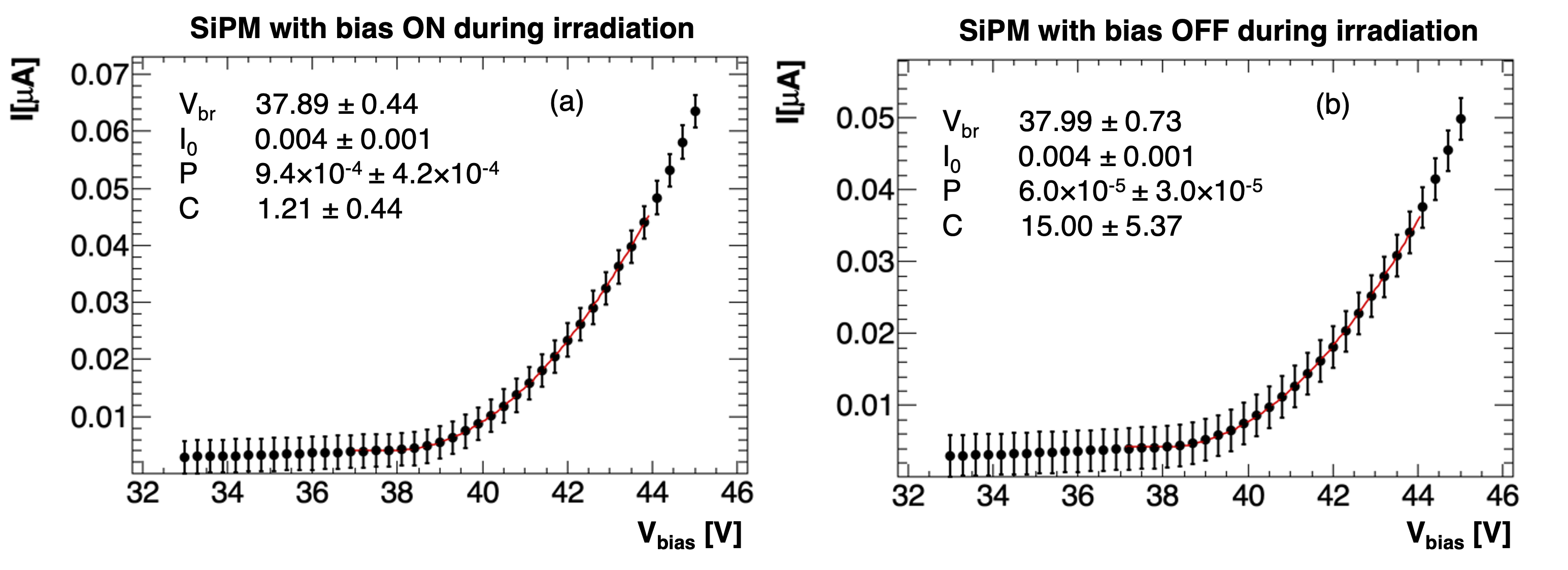}
    \caption{I-V curves of the two investigated SiPMs. One SiPM was biased (a) during the subsequent irradiation and the other was left un-biased (b). The values of parameters $V_{br}$, $I_0$, $P$, and $C$ obtained from the fit of the I-V curves are listed in the legends.} 
    \label{fig:IV_noIrr}
\end{figure}

Before the irradiation, the two SiPM samples were characterized by evaluating the I-V curves, as reported in Figure~\ref{fig:IV_noIrr}. During the following irradiation procedure, one of the two sensors was biased (Figure~\ref{fig:IV_noIrr}a) %DECIDERE COME CHIAMARLA a o left%
at $V_{op}$ while the other will be left unbiased (figure~\ref{fig:IV_noIrr}b). %DECIDERE COME CHIAMARLA b o right%.\\
Biasing the SiPM during irradiation can potentially reduce the supposed occurring TID effects, such as surface charge trapping. The applied electric field can influence the mobility and recombination of charge carriers, potentially mitigating the formation of interface states at the Si-SiO$_2$ boundary.

\begin{figure}[ht]
    \centering
    \includegraphics[width=\textwidth]{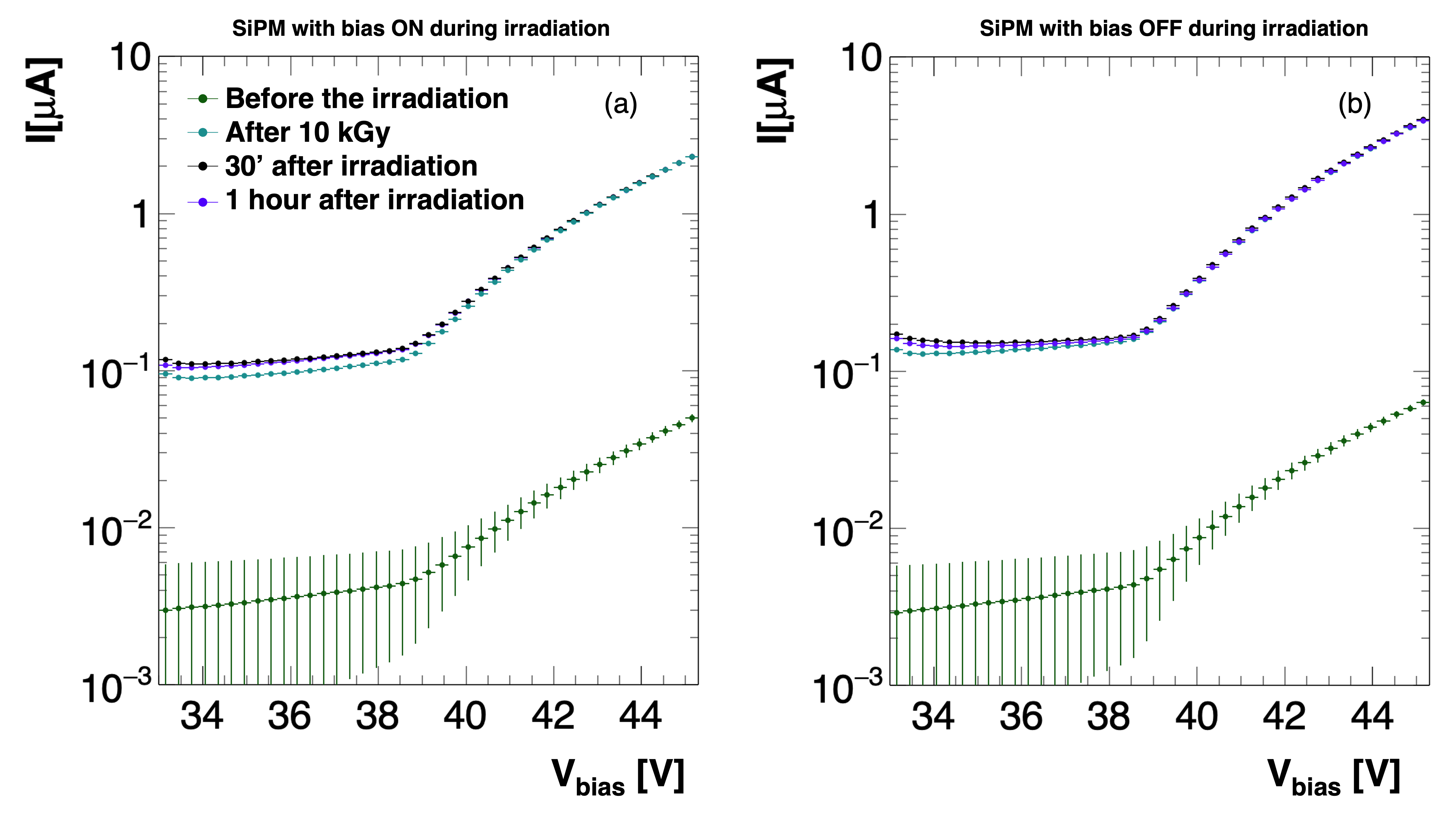}
    \caption{I-V curves at different steps of the test for SiPM kept biased (a) or unbiased (b) during the irradiation.} 
    \label{fig:IV_ONOFF} 
\end{figure}

Figure~\ref{fig:IV_ONOFF} shows the I-V curves acquired before the irradiation ($T_0$), after 10 kGy ($T_1$), 30 minutes after the irradiation ($T_2$), and 1 hour after the irradiation ($T_3$), for the two different SiPMs. As expected, the increase in the dark current is more pronounced in the case of the unbiased SiPM.

\begin{figure}[ht]
    \centering
    \includegraphics[width=\textwidth]{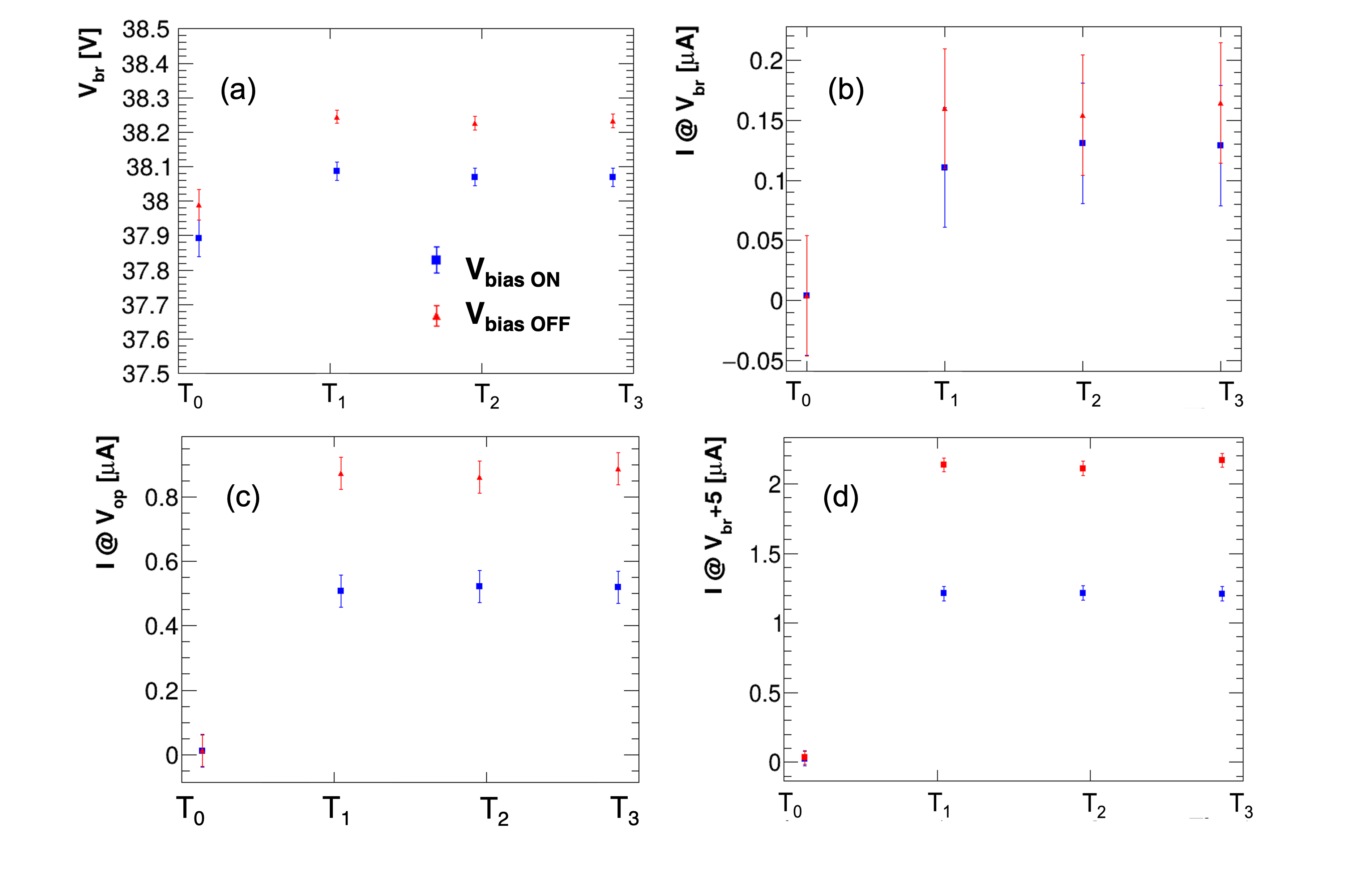}
    \caption{Breakdown voltage $V_{br} $(a), dark current at $V_{br}$ (b), dark current at operational voltage $V_{op}$ (c), and dark current evaluated at 5V over $V_{br}$ (d) for the biased (blue markers) and unbiased (red markers) components at different steps of irradiation. $T_0$, $T_1$, $T_2$ and $T_3$ correspond to the measurement time before, immediately after, 30 minutes and one hour after the  10 kGy irradiation, respectively.}
    \label{fig:casacciaSiPMRes} 
\end{figure}

From these curves, it was possible, following the previously described fitting procedure, to evaluate the $V_{br}$ and subsequently the operational voltage ($V_{op}=V_{br}+3$V)  and the corresponding dark current for both SiPMs at each irradiation step, as reported in figure~\ref{fig:casacciaSiPMRes}. 

Irradiation induces a negligible increment of breakdown voltage (figure~\ref{fig:casacciaSiPMRes}a) and a considerable increase, by nearly two orders of magnitude, of the dark current at different values of bias voltage (figure~\ref{fig:IV_ONOFF} and figure~\ref{fig:casacciaSiPMRes}, panels b, c, d). However, the SiPM that was biased during the irradiation presents a more limited increase of the breakdown voltage and of the dark current for $V_{bias}>V_{br}$ with respect to the unbiased one, confirming the protective effects of applied voltage on these components when exposed to radiation. One hour of quiescence does not produce any recovery of the radiation damage.

\section{Conclusions}
The CRILIN calorimeter, based on PbF$_2$ crystals and SiPMs, represents a promising alternative for high-energy particle detection in environments with intense radiation, such as the Muon Collider. This study aimed to evaluate the radiation resistance of the calorimeter's main components by subjecting the crystals and SiPMs to a gamma absorbed dose of roughly 10 kGy, simulating a 10-year service life.

The PbF$_2$ crystals exhibited a reduction in transmittance with partial recovery observed one hour after the irradiation. However, quantitative discrepancies between the two investigated samples in the radiation-induced loss of transparency and subsequent recovery indicates a need for improved production homogeneity. In contrast, the PWO-UF crystal demonstrated excellent radiation hardness, maintaining stable transmittance even at the highest absorbed doses.

The SiPMs also showed increased dark current following the irradiation, with a more pronounced effect in the unbiased sample.
Further research and optimization of materials and device configurations are already ongoing to enhance the reliability of the CRILIN calorimeter in high-radiation environments. In particular, the study of the effect of neutrons on crystals and SiPMs has already been planned and partially executed, along with a full characterization of the radiation effects on  full prototypes of the calorimeter, consisting of crystals optically coupled with SiPMs. The measurements on complete prototypes allow to investigate the optical coupling between crystals and SiPMs and the signal loss in the full detection chain induced by radiation. The results of this study will be the subject of a future publication.

\acknowledgments

This work was developed within the framework of the International Muon Collider Collaboration (\url{https://muoncollider.web.cern.ch}), where the Physics and Detector Group aims to evaluate potential detector R$\&$D to optimize experiment design in the multi-TeV energy regime. This work was supported by the EU Horizon 2020 Research and Innovation Programme under Grant Agreements No. 101006726 and No. 101004761. \\
The authors thank the LNF Division Research and Giuseppe Ferrara, ENEA NUC-IRAD-GAM Laboratory, for their technical and logistic support.

% Bibliography

%% [A] Recommended: using JHEP.bst file
%\bibliographystyle{JHEP}
%\bibliography{biblio.bib}

\begin{thebibliography}{99}
\bibitem{accettura2023towards}
C.~Accettura et~al., \emph{Towards a muon collider},
  \href{https://doi.org/10.1140/epjc/s10052-023-11889-x}{\emph{The European
  Physical Journal C} {\bfseries 83} (2023) 1}.

\bibitem{Cemmi_2022}
A.~Cemmi, A.~Colangeli, B.~D'Orsi, I.D.~Sarcina, E.~Diociaiuti, S.~Fiore
  et~al., \emph{{Radiation study of Lead Fluoride crystals}},
  \href{https://doi.org/10.1088/1748-0221/17/05/T05015}{\emph{Journal of
  Instrumentation} {\bfseries 17} (2022) T05015}.

\bibitem{Ceravolo_2022}
S.~Ceravolo, F.~Colao, C.~Curatolo, E.~Di~Meco, E.~Diociaiuti, D.~Lucchesi
  et~al., \emph{{Crilin: A CRystal calorImeter with Longitudinal InformatioN
  for a future Muon Collider}},
  \href{https://doi.org/https://doi.org/10.1088/1748-0221/17/09/p09033}{\emph{Journal
  of Instrumentation} {\bfseries 17} (2022) P09033}.

\bibitem{NIMA}
S.~Ceravolo, F.~Colao, C.~Curatolo, E.~{Di Meco}, E.~Diociaiuti, D.~Lucchesi
  et~al., \emph{Crilin: A crystal calorimeter with longitudinal information for
  a future muon collider},
  \href{https://doi.org/https://doi.org/10.1016/j.nima.2022.167817}{\emph{Nuclear
  Instruments and Methods in Physics Research Section A: Accelerators,
  Spectrometers, Detectors and Associated Equipment} {\bfseries 1047} (2023)
  167817}.

\bibitem{Frontiers}
C.~Cantone et~al., \emph{{Beam test, simulation, and performance evaluation of
  PbF2 and PWO-UF crystals with SiPM readout for a semi-homogeneous calorimeter
  prototype with longitudinal segmentation}},
  \href{https://doi.org/10.3389/fphy.2023.1223183}{\emph{Frontiers in Physics}
  {\bfseries 11} (2023) 1223183}.

\bibitem{BTF}
C.~Cantone, S.~Ceravolo, F.~Colao, E.~Di~Meco, E.~Diociaiuti, I.~Frank et~al.,
  \emph{{R\&D status for an innovative crystal calorimeter for the future Muon
  Collider}}, \href{https://doi.org/10.1109/TNS.2024.3364771}{\emph{IEEE
  Transactions on Nuclear Science} {\bfseries 71} (2024) 1116}.

\bibitem{KORZHIK_22}
M.~Korzhik, K.-T.~Brinkmann, V.~Dormenev, M.~Follin, J.~Houzvicka, D.~Kazlou
  et~al., \emph{{Ultrafast PWO scintillator for future high energy physics
  instrumentation}},
  \href{https://doi.org/https://doi.org/10.1016/j.nima.2022.166781}{\emph{Nuclear
  Instruments and Methods in Physics Research Section A: Accelerators,
  Spectrometers, Detectors and Associated Equipment} {\bfseries 1034} (2022)
  166781}.

\bibitem{Achenbach_98}
P.~Achenbach, I.~Altarev, K.~Grimm, T.~Hammel, D.~Von~Harrach, J.~Hoffmann
  et~al., \emph{{Radiation resistance and optical properties of lead fluoride
  Cherenkov crystals}},
  \href{https://doi.org/https://doi.org/10.1016/S0168-9002(98)00748-7}{\emph{Nuclear
  Instruments and Methods in Physics Research Section A: Accelerators,
  Spectrometers, Detectors and Associated Equipment} {\bfseries 416} (1998)
  357}.

\bibitem{Ren_GH_14}
G.-H.~Ren, X.-F.~Chen, H.-Y.~Li, Y.-T.~Wu, H.-S.~Shi and L.-S.~Qin,
  \emph{{Radiation Induced Optical Absorption of Cubic Lead Fluoride Crystals
  and the Effect of Annealing}},
  \href{https://doi.org/10.1088/0256-307X/31/8/086102}{\emph{Chinese Physics
  Letters} {\bfseries 31} (2014) 086102}.

\bibitem{Cantone_23a}
C.~Cantone, S.~Ceravolo, F.~Colao, E.~Di~Meco, E.~Diociaiuti, P.~Gianotti
  et~al., \emph{{R\&D status for an innovative crystal calorimeter for the
  future Muon Collider}},
  \href{https://doi.org/https://doi.org/10.1051/epjconf/202328802002}{\emph{EPJ
  Web of Conferences} {\bfseries 288} (2023) 02002}.

\bibitem{GARUTTI201969}
E.~Garutti and Y.~Musienko, \emph{{Radiation damage of SiPMs}},
  \href{https://doi.org/https://doi.org/10.1016/j.nima.2018.10.191}{\emph{Nuclear
  Instruments and Methods in Physics Research Section A: Accelerators,
  Spectrometers, Detectors and Associated Equipment} {\bfseries 926} (2019)
  69}.

\bibitem{HamamatsuData}
\url{https://www.hamamatsu.com/content/dam/hamamatsu-photonics/sites/documents/99_SALES_LIBRARY/ssd/s14160-1310ps_etc_kapd1070e.pdf}.

\bibitem{Fienberg_15}
A.T.~Fienberg et~al., \emph{{Studies of an array of PbF$_2$ Cherenkov crystals
  with large-area SiPM readout}},
  \href{https://doi.org/https://doi.org/10.1016/j.nima.2015.02.028}{\emph{Nuclear
  Instruments and Methods in Physics Research Section A: Accelerators,
  Spectrometers, Detectors and Associated Equipment} {\bfseries 783} (2015)
  12}.

\bibitem{Frankenthal_19}
A.~Frankenthal et~al., \emph{{Characterization and performance of PADME’s
  Cherenkov-based small-angle calorimeter}},
  \href{https://doi.org/https://doi.org/10.1016/j.nima.2018.12.035}{\emph{Nuclear
  Instruments and Methods in Physics Research Section A: Accelerators,
  Spectrometers, Detectors and Associated Equipment} {\bfseries 919} (2019)
  89}.

\bibitem{Calliope_report}
S.~Baccaro, A.~Cemmi, I.~Di~Sarcina and G.~Ferrara, \emph{{Gamma Irradiation
  CALLIOPE facility at ENEA Casaccia Research Centre}},  Tech. Rep.
  \href{https://iris.enea.it/bitstream/20.500.12079/6838/1/RT-2019-04-ENEA.pdf}{RT/2019/4},
  ENEA, Rome, Italy (2019).

\bibitem{ISO_51607_12}
\emph{{Practice for use of the alanine-EPR dosimetry system}},  Tech. Rep.
  \href{https://www.iso.org/standard/62955.html}{ISO/ASTM DID 51607},
  International Organization for Standardization, Geneva, CH (2012).

\bibitem{Rodnyi1997}
P.~Rodnyi, \emph{{Physical Processes in Inorganic Scintillators}}, Laser and
  Optical Science and Technology Series. CRC Press, Boca Raton, FL (1997).

\bibitem{Weber2004}
M.~Weber, \emph{Scintillation: mechanisms and new crystals},
  \href{https://doi.org/https://doi.org/10.1016/j.nima.2004.03.009}{\emph{Nuclear
  Instruments and Methods in Physics Research Section A: Accelerators,
  Spectrometers, Detectors and Associated Equipment} {\bfseries 527} (2004) 9}.

\bibitem{Potylitsyn2021}
A.~Potylitsyn, G.~Kube, A.~Novokshonov, A.~Vukolov, S.~Gogolev, B.~Alexeev
  et~al., \emph{First observation of quasi–monochromatic optical cherenkov
  radiation in a dispersive medium (quartz)},
  \href{https://doi.org/https://doi.org/10.1016/j.physleta.2021.127680}{\emph{Physics
  Letters A} {\bfseries 417} (2021) 127680}.

\bibitem{Lin2001}
Q.~Lin, X.~Feng, Z.~Man, Y.~Zhang, Z.~Yin and Q.~Zhang, \emph{Origin of the
  radiation-induced 420nm color center absorption band in {P}b{WO}$_{4}$
  crystal},
  \href{https://doi.org/https://doi.org/10.1016/S0038-1098(01)00097-7}{\emph{Solid
  State Communications} {\bfseries 118} (2001) 221}.

\bibitem{Deng1999}
Q.~Deng, Z.~Yin and R.-Y.~Zhu, \emph{Radiation-induced color centers in
  la-doped {P}b{WO}$_{4}$ crystals},
  \href{https://doi.org/10.1016/S0168-9002(99)00835-9}{\emph{Nuclear
  Instruments and Methods in Physics Research Section A-accelerators
  Spectrometers Detectors and Associated Equipment} {\bfseries 438} (1999)
  415}.

\bibitem{Wensheng2000}
W.~Li, T.~Tang and X.~Feng, \emph{{Complex color centers in ultraviolet-light
  irradiated PbWO4 single crystal}},
  \href{https://doi.org/10.1063/1.373442}{\emph{Journal of Applied Physics}
  {\bfseries 87} (2000) 7692}.

\bibitem{Baccaro2015}
S.~Baccaro, A.~Cemmi, I.~Di~Sarcina and F.~Menchini, \emph{{Gamma Rays Effects
  on the Optical Properties of Cerium-Doped Glasses}},
  \href{https://doi.org/https://doi.org/10.1111/ijag.12131}{\emph{International
  Journal of Applied Glass Science} {\bfseries 6} (2015) 295}.

\bibitem{Baccaro2001}
S.~Baccaro, A.~Cecilia, A.~Cemmi, E.~Mihokova, M.~Nikl, K.~Nitsch et~al.,
  \emph{Colour centres induced by $\gamma$ irradiation in scintillating glassy
  matrices for middle and low energy physics experiments},
  \href{https://doi.org/https://doi.org/10.1016/S0168-583X(01)00760-1}{\emph{Nuclear
  Instruments and Methods in Physics Research Section B: Beam Interactions with
  Materials and Atoms} {\bfseries 185} (2001) 294}.

\bibitem{Addesa2020}
F.~Addesa, M.~Campana, A.~Cemmi, B.~D'Orsi, I.~Dafinei, I.~Di~Sarcina et~al.,
  \emph{{Optical spectroscopic characterization of LYSO crystals at the
  Calliope facility (ENEA Casaccia R.C.)}},  Tech. Rep.
  \href{https://iris.enea.it/retrieve/dd11e37c-ead0-5d97-e053-d805fe0a6f04/RT-2020-15-ENEA.pdf}{RT/2020/15},
  ENEA, Rome, Italy (2020).

\bibitem{LINDSTROM}
G.~Lindström, \emph{{Radiation damage in silicon detectors}},
  \href{https://doi.org/https://doi.org/10.1016/S0168-9002(03)01874-6}{\emph{Nuclear
  Instruments and Methods in Physics Research Section A: Accelerators,
  Spectrometers, Detectors and Associated Equipment} {\bfseries 512} (2003)
  30}.

\bibitem{PicoAmmeter}
\url{https://www.testequipmenthq.com/datasheets/KEITHLEY-6487-Datasheet.pdf}.

\bibitem{para}
A.~Nagai, N.~Dinu and A.~Para, \emph{{Breakdown voltage and triggering
  probability of SiPM from IV curves}},  in \emph{2015 IEEE Nuclear Science
  Symposium and Medical Imaging Conference (NSS/MIC)}, pp.~1--4, IEEE, 2015.
\end{thebibliography}

%% or
%% [B] Manual formatting (see below)
%% (i) We suggest to always provide author, title, and journal data or doi:
%% in short all the information that clearly identify a document.
%% (ii) Please avoid comments such as "For a review", "For some examples",
%% "and references therein" or move them in the text. In general, please leave only references in the bibliography and move all accessory text to footnotes.
%% (iii) Also, please have only one work for each \bibitem.

\end{document}